\documentclass[twocolumn,showpacs,amsmath,amssymb,eqsecnum,nofootinbib]{revtex4}

\begin{document}
\title{Electromagnetic Field in Higher-Dimensional Black-Hole Spacetimes\footnote{%
The results contained in this paper have been presented at 
the Black hole VI conference in White Point, Canada, May 12--16 2007, and at 
the GRG18 conference
in Sydney, Australia, July 8--13 2007.}}

\author{Pavel Krtou\v{s}}

\email{Pavel.Krtous@mff.cuni.cz}

\affiliation{Institute of Theoretical Physics, Faculty of Mathematics and Physics, Charles University in Prague,\\
V~Hole\v{s}ovi\v{c}k\'ach 2, Prague, Czech Republic}

\date{June 28, 2007}  

\begin{abstract}
A special test electromagnetic field in the spacetime of the 
higher-dimensional generally rotating NUT--(A)dS black hole is found. 
It is adjusted to the hidden symmetries of the background 
represented by the principal Killing-Yano 
tensor. Such electromagnetic field generalizes the field of 
charged black hole in four dimensions.  
In higher dimensions, however, the gravitational back reaction
of such a field cannot be consistently solved.
\end{abstract}


\vspace*{4ex}

\maketitle

\section{Introduction}
\label{sc:intro}

In recent years people have become interested 
in higher-dimensional spacetimes with motivation, 
among others, coming from string theories. One of the 
important classes of exact solutions
of the Einstein equations are the spacetimes 
representing black holes. 
In the last years various generalizations of black hole solutions 
to the higher-dimensional gravity have been discovered.
Starting with nonrotating and rotating black holes 
\cite{Tangherlini:1963,MyersPerry:1986} and
following with the metric describing the rotating black hole 
with the cosmological constant in five dimensions \cite{HawkingEtal:1999},
the generally rotating black hole 
in arbitrary dimension with
the cosmological constant was found in 
\cite{GibbonsEtal:2004,GibbonsEtal:2005}.
In 2006 this solution was rewritten \cite{ChenLuPope:2006} in much more 
convenient coordinates which allowed to add 
NUT parameters and compute 
the curvature tensors explicitly~\cite{HamamotoEtal:2007}.

Recently, the properties of this solution have been discussed
in various papers. It has been shown that the spacetime
possesses hidden symmetries which can be described by
the principal Killing-Yano tensor \cite{KubiznakFrolov:2007},
the properties of which have been thoroughly discussed in
\cite{KrtousEtal:2007a}. One of the main consequences
of these hidden symmetries is a possibility to
find a complete set of integration constants for 
a geodesic motion which are in involution 
\cite{PageEtal:2007,KrtousEtal:2007b,KrtousEtal:2007a}. This result is
related to the the separability of the Hamilton-Jacobi
equation which, together with the separability 
of the Klein-Gordon equation, was demonstrated in~\cite{FrolovEtal:2007}.

It seems that a generalization of the 
black hole solution to include an electromagnetic 
field is not straightforward. 
Some partial results are know 
\cite{AlievFrolov:2004,Aliev:2005,Aliev:2006,Aliev:2007,BrihayeDelsate:2007,KunzEtal:2007},
however a solution representing
a rotating black hole with electromagnetic field 
parameterized by full number of black hole parameters 
and electric/magnetic charges is not known.

In this work we present the test electromagnetic field
on the background described by the metric
with the structure of black hole solution
\cite{ChenLuPope:2006}, however,
we do not enforce the specific form of the metric function
${X_\mu}$ which is needed to satisfy the vacuum Einstein
equations. The electromagnetic field is
described by ${n=\lfloor D/2\rfloor}$ electric and
magnetic charges and it is adjusted to the
explicit as well as hidden symmetries of the spacetime. 
In even dimensions our field is equivalent to the harmonic forms
independently found in \cite{ChenLu:2007}, 
our results supply the proof that 
these forms are harmonic in all even dimensions.

We also show that our field is a generalization
of the electromagnetic field known
from the Carter \cite{Carter:1968} and 
Pleba\'nski--Demia\'nski \cite{PlebanskiDemianski:1976} form 
of the black hole solution 
in ${D=4}$. 
In this special case, however, the metric function
can be modified in such a way that that 
the metric and the electromagnetic field satisfy
the full coupled Einstein--Maxwell equations.
Unfortunately, the same is not true in the higher-dimensional case.

The plan of the paper is following. We discuss separately
even and odd spacetime dimensions---although our
procedure is very similar in both cases, the resulting
expressions are slightly different and it would
be artificial to try to \vague{squeeze} them
into the same equations. We discuss the 
case of even dimensions in detail and than sketch 
the corresponding results for the odd dimensions.
First we review the metric and the symmetries of
the black hole spacetime, next we find the test electromagnetic field
\vague{adjusted} to the symmetries of the spacetime.
For even dimensions we also discuss the case of
the physical dimension ${D=4}$, in the odd dimension case we study
the Chern--Simons modification of the electromagnetic field. 
Finally, we discuss a possibility to generalize 
our fields to a solution of the coupled Einstein--Maxwell equations.
The results are recapitulated in the Summary and the paper
is concluded by the technical Appendix.

\vspace*{-1ex}

\section{Even dimensions, ${D=2n}$}
\vspace*{-1.5ex}

\subsection*{The metric}

We start with a simpler case of even spacetime dimension ${D=2n}$.
The spacetime of generally rotating black hole with NUT charges
can be described by the metric
\begin{equation}\label{metricE}
\tens{g}=\sum_{\mu=1}^n\;\biggl[\; \frac{U_\mu}{X_\mu}\,{\grad x_{\mu}^{\;\,2}}
  +\, \frac{X_\mu}{U_\mu}\,\Bigl(\,\sum_{k=0}^{n-1} \A{k}_{\mu}\grad\psi_k \Bigr)^{\!2} \;\biggr]
  \period
\end{equation}
Here ${x_\mu}$, ${\mu=1,\dots,n}$, correspond to radial\footnote{%
The radial coordinate (and some other related quantities)
are actually rescaled by the imaginary unit ${i}$ 
in order to put the metric to a more symmetric and compact form
--- cf., e.g., \cite{ChenLuPope:2006}.}
and latitudinal directions, while ${\psi_j}$, ${j=0,\dots,n-1}$,
to temporal and longitudinal directions. The functions
${U_\mu}$ and ${\A{k}_\mu}$ (together with ${U}$ and ${\A{k}}$ used below) are defined as
\begin{gather}\label{UAdef}
  U_{\mu}=\prod_{\substack{\nu=1\\\nu\ne\mu}}^{n}(x_{\nu}^2-x_{\mu}^2)\comma
  U=\prod_{\substack{\mu,\nu=1\\\mu<\nu}}^n (x_\mu^2-x_\nu^2)\commae\\
  \A{k}_{\mu}=\!\!\!\!\!\sum_{\substack{\nu_1,\dots,\nu_k=1\\\nu_1<\dots<\nu_k,\;\nu_i\ne\mu}}^n\!\!\!\!\!x^2_{\nu_1}\dots x^2_{\nu_k}\comma
  \A{k}=\!\!\!\!\!\sum_{\substack{\nu_1,\dots,\nu_k=1\\\nu_1<\dots<\nu_k}}^n\!\!\!\!\!x^2_{\nu_1}\dots x^2_{\nu_k}\period\raisetag{3ex}
\end{gather}
We call the remaining functions ${X_\mu}$ \defterm{the metric functions}.

In the following we consider a broader class of spacetimes than merely
the black hole spacetime, namely we just assume that the metric functions ${X_\mu}$ 
depend arbitrarily on a \emph{single} coordinate, ${X_\mu=X_\mu(x_\mu)}$. For the
black hole spacetime they acquire a specific from (see \eqref{BHXsE} below) 
determined by the vacuum Einstein equations. 

The metric can be diagonalized
\begin{equation}
\tens{g}
     = \sum_{\mu=1}^n\,
          \biggl(\,\frac{U_\mu}{X_\mu}\,\ep^\mu \ep^\mu 
          + \frac{X_\mu}{U_\mu}\,\ep^{\hat \mu} \ep^{\hat \mu}\,\biggr)
     = \sum_{a=1}^D \eb^a \eb^a
\end{equation}
introducing the unnormalized frame ${\ep^a}$ and normalized frame ${\eb^a}$ of 1-forms 
\begin{equation}\label{eforms}
\begin{aligned}
\ep^\mu &= \grad x_{\mu}\commae&
\eb^{\mu} &= {\textstyle\bigl[\frac{U_\mu}{X_\mu}\bigr]^{\!1\!/2}}\,\ep^{\mu}\commae\\
\ep^{\hat\mu} &= \sum_{k=0}^{n-1}\A{k}_{\mu}\grad\psi_k\commae&
\eb^{\hat\mu} &= {\textstyle\bigl[\frac{X_\mu}{U_\mu}\bigr]^{\!1\!/2}}\,\ep^{\hat\mu}\period
\end{aligned}
\end{equation}
The dual frames of vectors are given by 
\begin{equation}\label{evectors}
\begin{aligned}
\ep_\mu &= \cv{x_\mu}\commae&
\eb_{\mu} &= {\textstyle\bigl[\frac{X_\mu}{U_\mu}\bigr]^{\!1\!/2}}\,\ep_{\mu}\commae\\
\ep_{\hat\mu}&= \sum_{k=0}^{n-1}\frac{(-x_{\mu}^2)^{n\!-\!1\!-\!k}}{U_{\mu}}\,\cv{\psi_{k}}\commae&
\eb_{\hat\mu} &= {\textstyle\bigl[\frac{U_\mu}{X_\mu}\bigr]^{\!1\!/2}}\,\ep_{\hat\mu}\period
\end{aligned}
\end{equation}
Here we use the convention ${\hat\mu=\mu+n}$ and Greek indieces run from ${1}$ to ${n}$.
The inverse relations can be easily obtained with help of the relations \eqref{AUrel} in the Appendix.

Let us note that the definition of unnormalized 1-forms ${\ep^a}$ and vectors ${\ep_a}$
does not depend on the metric functions~${X_\mu}$.

It was shown in \cite{HamamotoEtal:2007} that the Ricci curvature 
for such a metric is
\begin{equation}\label{RicciE}
  \Ric=-\sum_{\mu=1}^n \; r_\mu\; \bigl(\eb^{\mu}\eb^{\mu}+\eb^{\hat\mu}\eb^{\hat\mu}\bigl)\commae
\end{equation}
where 
\begin{equation}
\begin{split}
  r_\mu 
    &= \frac12\,\frac{X_\mu''}{U_\mu}
      +\sum_{\substack{\nu=1\\\nu\neq\mu}}^n\frac1{U_\nu}\,\frac{x_\nu X_\nu' \!-\! x_\mu X_\mu'}{x_\nu^2\!-\!x_\mu^2}
      -\sum_{\substack{\nu=1\\\nu\neq\mu}}^n\frac1{U_\nu}\,\frac{X_\nu \!-\! X_\mu}{x_\nu^2\!-\!x_\mu^2}\\
    &= \frac\partial{\partial x_\mu^2}\Biggl[\,\sum_{\nu=1}^{n}\frac{x_\nu^2\bigl(x_\nu^{-1}X_\nu\bigr)_{\!,\nu}}{U_\nu}\Biggr]\period
\end{split}\raisetag{5ex}
\end{equation}
The scalar curvature simplifies to
\begin{equation}\label{sccurE}
  \sccur = -\sum_{\nu=1}^n \; \frac{X_\nu''}{U_\nu}\period
\end{equation}
Here, the primes denote the differentiation with respect to the single argument
of the metric function, ${X_\mu'=X_{\mu,\mu}}$.

The requirement that the metric \eqref{metricE} satisfies the Einstein equations
with the cosmological constant implies that the metric functions have the form
\begin{equation}\label{BHXsE}
  X_\mu = b_\mu\, x_\mu + \sum_{k=0}^{n}\, c_{k}\, x_\mu^{2k}\commae
\end{equation}
see \cite{ChenLuPope:2006,HamamotoEtal:2007} or \eqref{sumhU} in the Appendix.
The constants ${c_k}$ and ${b_\mu}$ are then related to the cosmological constants,
angular momenta, mass, and NUT charges, see, e.g., \cite{ChenLuPope:2006}
for details. For ${b_\mu=0}$ we obtain \cite{HamamotoEtal:2007} 
the constant curvature spacetime with
the scalar curvature ${\sccur=-2n(2n\!-\!1)c_{2n}}$. 

The spacetime with the metric \eqref{metricE} possesses the explicit
symmetries given by the Killing vectors ${\cv{\psi_j}}$ and hidden 
symmetries which are related to the principal Killing-Yano tensor
discovered in \cite{KubiznakFrolov:2007} and discussed in detail 
in \cite{KrtousEtal:2007a}. The principal Killing-Yano tensor is dual
to the rank-2 closed conformal Killing-Yano tensor 
which has a very simple form in the frames ${\ep^a}$ and ${\eb^a}$:
\begin{equation}\label{CCKY}
\tens{h} 
 = \sum_{\mu=1}^n \; x_\mu\; \ep^\mu \wedge \ep^{\hat\mu}
 = \sum_{\mu=1}^n \; x_\mu\; \eb^\mu \wedge \eb^{\hat\mu}\period
\end{equation}
It has been demonstrated in \cite{KrtousEtal:2007a} that 
it is possible to generate a series of higher-rank Killing--Yano tensors 
and a series of rank-2 Killing tensors from ${\tens{h}}$. 
The conformal Killing--Yano tensor also identifies eigenspaces spanned on the pairs
${\{\eb_\mu,\,\eb_{\hat\mu}\}}$ and coordinates ${x_\mu}$ as corresponding eigenvalues.

\subsection*{Algebraically special test electromagnetic field}

Now we turn to the task to find an algebraically special test electromagnetic fields
on the background given by the metric \eqref{metricE}.
By \emph{algebraically special} we mean that the Maxwell tensor ${\tens{F}}$
shares the explicit symmetry of the metric (it is independent of ${\psi_j}$)
and it is aligned with the hidden symmetry of the spacetime, namely it
has the same eigenspaces as the principal conformal Killing-Yano tensor ${\tens{h}}$.
We thus require 
\begin{equation}\label{EMFansatz}
  \tens{F} = \sum_{\mu=1}^n\, f_\mu\; \ep^\mu\wedge\ep^{\hat\mu}\comma 
  f_\mu=f_\mu(x_1,\dots,x_n)\period
\end{equation}

The Maxwell tensor is generated by the vector potential,
${\tens{F}=\grad\tens{A}}$. We assume the vector potential
\begin{equation}\label{EMAansatz}
  \tens{A} = \sum_{\mu=1}^n \bigl(A_{\mu}\,\ep^{\mu}+A_{\hat\mu}\,\ep^{\hat\mu}\bigr)
\end{equation}
with the components ${A_\mu}$ and ${A_{\hat\mu}}$ independent of ${\psi_j}$.
Comparing ${\grad\tens{A}}$ with \eqref{EMFansatz} we find that ${A_\mu}$
terms are gauge-trivial and ${A_{\hat\mu}}$ must satisfy 
${\bigl((x_\nu^2-x_\mu^2)A_{\hat\mu}\bigr)_{\!,\nu}=0}$,
from which follows that the vector potential can be written as
\begin{equation}\label{EMAE}
  \tens{A} = \sum_{\mu=1}^n \frac{g_\mu\, x_\mu}{U_\mu}\;\ep^{\hat\mu}\commae
\end{equation}
where ${g_\mu}$ are functions of a \emph{single variable} only,
${g_\mu = g_\mu(x_\mu)}$. Evaluating the Maxwell tensor we get the components ${f_\mu}$:
\begin{equation}\label{EMfE}
  f_\mu =  \frac{g_\mu}{U_\mu}+ \frac{x_\mu\,g_\mu'}{U_\mu}
  +2\,x_\mu\,\sum_{\substack{\nu=1\\\nu\neq\mu}}^n\frac{1}{U_\nu}\,\frac{x_\nu\,g_\nu - x_\mu\,g_\mu}{x_\nu^2-x_\mu^2}\period
\end{equation}

Alternatively, we could apply directly the first Maxwell equation ${\grad\tens{F}=0}$
to the Maxwell tensor \eqref{EMFansatz}. With help of identity \eqref{deps} we find
that ${f_\mu}$ are generated by an auxiliary potential ${\phi}$, 
\begin{equation}\label{EMfphi}
  f_\mu = \phi_{,\mu}\commae
\end{equation}
which satisfies the equation 
\begin{equation}\label{EMphieq}
    \phi_{,\mu\nu} = 2\; \frac{x_\nu\,\phi_{,\mu}-x_\mu\,\phi_{,\nu}}{x_\mu^2-x_\nu^2}\qquad\text{for}\;\mu\neq\nu\period
\end{equation}
The field \eqref{EMAE} found above is generated by the potential
\begin{equation}\label{EMphi}
    \phi = \sum_{\nu=1}^n \frac{g_\nu\, x_\nu}{U_\nu}\period
\end{equation}

Next we proceed to calculate the source ${\tens{J}}$ of the electromagnetic field 
using the second Maxwell equation ${\tens{J}=-\covd\cdot\tens{F}}$. Expressing 
the Maxwell tensor in coordinates ${x_\mu,\psi_j}$, using 
${F^{na}{}_{;n}=\detg^{-1/2}\sum_\nu(\detg^{1/2}F^{\nu a})_{\!,\nu}}$,
the fact that the determinant ${\detg}$ of the metric in these coordinates 
is ${\detg=U^2}$, identities \eqref{AUrel}, \eqref{sumn1jAxn1j},
and relations \eqref{EMFansatz}, \eqref{EMfphi} 
we obtain
\begin{equation}\label{EMJ}
  \tens{J} = \sum_{\mu=1}^n j_\mu\;\ep_{\hat\mu}\commae
\end{equation}
with
\begin{equation}\label{EMjphi}
  j_\mu = -2\frac\partial{\partial x_\mu^2}\biggl[\phi-x_\mu^2\sum_{\nu=1}^n x_\nu^{-1}\,\phi_{,\nu}\biggr]\period
\end{equation}
Substituting \eqref{EMphi} we finally obtain
\begin{equation}\label{EMj}
  j_\mu = 2\frac\partial{\partial x_\mu^2}\Biggl[\sum_{\nu=1}^n\frac{x_\nu^2\,g_{\nu}'}{U_\nu}\Biggr]\period
\end{equation}
It is worth to mention that the expression for the source~${\tens{J}}$
does not contain any reference to the metric functions~${X_\mu}$.

We are interested in the electromagnetic field without sources, so we require
${\tens{J}=0}$. Integrating \eqref{EMj} we find that the sum in the square brackets 
has to be a constant. However, this sum has a special form discussed in the Appendix.
Using \eqref{sumhU} we find that ${g_\mu'}$ are given by a single polynomial of 
the ${(n\!\!-\!\!1)}$-th order in variable ${x_\mu^2}$. Integrating once more we find
\begin{equation}\label{EMggenE}
  g_\mu\, x_\mu =  e_\mu x_\mu + \sum_{k=0}^{n-1}\, a_k\, \bigl(-x_\mu^2\bigr)^{n-1-k}\period
\end{equation}
Substituting into the vector potential \eqref{EMAE} or the scalar potential \eqref{EMphi}
we find with help of the relations \eqref{AUrel} that the terms containing the constants
${a_k}$ are gauge trivial (they contribute by ${\sum_{k=0}^{n-1}a_k\grad\psi_k}$ into ${\tens{A}}$
or by just the constant ${a_0}$ into ${\phi}$) and they can be ignored. 

We thus have found that the algebraically special electromagnetic field
(i.e., the field of the form \eqref{EMFansatz}) satisfies the Maxwell equations
on the background described by the metric \eqref{metricE} if and only if
it is generated by the vector potential 
\begin{equation}\label{EMAres}
  \tens{A} =  \sum_{\mu=1}^n \frac{e_\mu\,x_\mu}{U_\mu}\, \ep^{\hat\mu}\period
\end{equation}
The components ${f_\mu}$ of the Maxwell tensor are easily determined by \eqref{EMfphi} 
from the auxiliary potential
\begin{equation}\label{EMphires}
  \phi = \sum_{\mu=1}^n \frac{e_\mu\,x_\mu}{U_\mu}\commae
\end{equation}
and they are
\begin{equation}\label{EMfres}
  f_\mu = \frac{e_\mu}{U_\mu}
  +2\,x_\mu\,\sum_{\substack{\nu=1\\\nu\neq\mu}}^n\frac{1}{U_\nu}\,\frac{x_\nu\,e_\nu - x_\mu\,e_\mu}{x_\nu^2-x_\mu^2}\period
\end{equation}
Here, ${e_\mu}$ are constants which can be related
using the Gauss and Stokes theorems to the electric and magnetic charges of the field. 

If we set all charges except one, say ${e_\nu}$, to zero, the 
Maxwell tensor ${\tens{F}}$ corresponds to the harmonic form 
${\tens{G}^{(\nu)}_{(2)}}$ recently found and verified 
for particular cases in \cite{ChenLu:2007}.

The surprising property of our field is that it satisfies the Maxwell equations
independently of a specific form of the metric functions ${X_\mu}$. 
Moreover, as we will see in \eqref{EMT},
the stress-energy tensor corresponding to the field \eqref{EMFansatz}
has the form consistent with the structure of the Ricci (and the Einstein) 
tensor \eqref{RicciE}.
These facts open a possibility that we could solve 
the full Einstein--Maxwell equations: 
modifying the metric functions ${X_\mu}$ we could construct the 
spacetime in which the stress-energy tensor~${\tens{T}}$ 
would be a source for the Einstein equations,
and the electromagnetic field would remain the 
solution of the Maxwell equations. 

First, we will show that this goal can be achieved in ${D=4}$
dimensions. Unfortunately, next we will demonstrate that 
this procedure does not work in higher dimensions.

\subsection*{Case ${D=4}$}

The metric \eqref{metricE}  
is a generalization of the ${D=4}$ black hole solution 
in the form found by Carter \cite{Carter:1968} and elaborated 
by Pleba\'nski and Demia\'nski \cite{PlebanskiDemianski:1976}.
The full class of Pleba\'nski--Demia\'nski solutions includes
accelerated rotating NUT and electromagnetically charged black holes. 
The meaning of all parameters of the solutions was recently discussed in the series of papers 
\cite{GriffithsPodolsky:2005,GriffithsPodolsky:2006a,GriffithsPodolsky:2006b,PodolskyGriffiths:2006,GriffithsPodolsky:2007}. 
We use the charged 
Pleba\'nski--Demia\'nski metric with the acceleration set to zero:\footnote{%
See, e.g., eqs.~(5) and (6) of \cite{GriffithsPodolsky:2006b} with acceleration ${\alpha=0}$. 
Here we also used the gauge freedom to set ${\omega=1}$
and redefined ${k}$.}
\begin{equation}\label{P-D}
\begin{split}
\tens{g} &= -\frac{Q}{r^2+p^2}\,\bigl(\grad t +p^2 \grad\sigma\bigr)^2+\frac{r^2+p^2}{Q}\,\grad r^2\\
         &\quad+\frac{r^2+p^2}{P}\,\grad p^2+\frac{P}{r^2+p^2}\,\bigl(\grad t -r^2 \grad\sigma\bigr)^2\commae
\end{split}
\end{equation}
where
\begin{equation}\label{QPfc}
\begin{aligned}
Q &= e^2\!-\! 2mr \!+\! k \!+\!\eps r^2 \!+\!\lambda r^4  = \mathcal{X}(-r^2)-2mr+e^2\commae\\
P &= -g^2 \!+\! 2np \!+\! k \!-\!\eps p^2 \!+\!\lambda p^4 = \mathcal{X}(p^2)+2np-g^2\period
\end{aligned}
\end{equation}
The electromagnetic field is given by
\begin{equation}\label{P-DEMA}
\tens{A} = -\frac{1}{r^2+p^2}\;
               \Bigl( e\,r\,\bigl(\grad t + p^2\,\grad\sigma\bigr)
               +g\,p\,\bigl(\grad t - r^2\,\grad\sigma\bigr) \Bigr)\period
\end{equation}
Here ${t,r,p,\sigma}$ are temporal, radial, latitudinal, and longitudinal coordinates,
${m,n,\eps,\lambda}$ are parameters related to mass, NUT charge, angular momentum, and cosmological constant,
and ${e,g}$ are electric and magnetic charges. 

This metric and the electromagnetic field satisfy coupled Einstein--Maxwell equations.
For ${e,g=0}$ the metric \eqref{P-D} satisfies the vacuum Einstein equations,
however, even in this case the electromagnetic field \eqref{P-DEMA} (with non-zero charges)
satisfies the Maxwell equations and it is thus a valid test electromagnetic field on the 
given vacuum background.

Now, we can easily identify the Pleba\'nski--Demia\'nski ${\tens{g}}$ and ${\tens{A}}$ 
with our general metric \eqref{metricE}
and field \eqref{EMAres} using the dictionary 
\begin{equation}
\begin{aligned}
  \psi_0 &=t\,,\!\!\!      &  x_1 &= ir\,,\!\!\! &  X_1 &= Q\,,\!\!\!  &  U_1 &= r^2+p^2\,,\!\!\!    & e_1 &= ie\commae\\
  \psi_1 &= \sigma\,,\!\!\!&  x_2 &= p\,,\!\!\!  &  X_2 &= P\,,\!\!\!  &  U_2 &= -r^2-p^2\,,\!\!\! & e_2 &= g\period 
\end{aligned}
\end{equation}
Clearly, if the charges ${e,g}$ are missing in the metric functions ${Q}$ and ${P}$,
these functions corresponds exactly to ${X_1}$ and ${X_2}$ given by \eqref{BHXsE}
and the electromagnetic field \eqref{P-DEMA} corresponds to the test field given by \eqref{EMAres}.
However, it is possible to modify ${P}$ and ${Q}$ 
by adding ${e^2}$ and ${-g^2}$ respectively
(i.e., changing the vacuum ${X_\mu}$\!'s  by adding ${-e_\mu^2}$), 
and we obtain the metric and the electromagnetic field satisfying the 
coupled Einstein--Maxwell equations.

\subsection*{Einstein--Maxwell equations in even ${D\neq4}$}

In a generic dimension we first evaluate the stress-energy tensor ${\tens{T}}$
of the electromagnetic field \eqref{EMFansatz}. A straightforward calculation leads to
\begin{equation}\label{EMT}
8\pi\,\tens{T} 
   = \sum_{\mu=1}^n\; \bigl( 2 f_\mu^2 - f^2\bigr)\;
        \bigl(\eb^\mu\eb^\mu+\eb^{\hat\mu}\eb^{\hat\mu}\bigr)\commae
\end{equation}
with the trace 
\begin{equation}\label{EMTrT}
  8\pi\,T = 2\,(2-n)\; f^2\commae
\end{equation}
where the function ${f^2}$ is defined as
\begin{equation}\label{EMf2}
  f^2 = \sum_{\nu=1}^n f_\nu^2\period
\end{equation}
We explicitly see that the trace of the stress-energy is non-vanishing 
for ${D\neq4}$ which is related to the fact that the electromagnetic field 
is not conformally invariant in a general dimension.

Now we would like to solve the Einstein equations
${\Ric - \frac12 \sccur \tens{g} + \Lambda \tens{g} = 8\pi\,\tens{T}}$.
The trace gives the condition
\begin{equation}\label{sccurcond}
  \sccur = 2\frac{D}{D-2}\;\Lambda + 2\,\frac{D-4}{D-2} \;f^2\period
\end{equation}
However, the scalar curvature has the form \eqref{sccurE}
and it immediately follows that
\begin{equation}\label{sccurUder}
    \frac{\partial^{2n-2}}{\partial x_\mu^{2n-2}}\bigl(U_\mu\,\sccur\bigr) = - X_\mu^{[2n]}\commae
\end{equation}
which is a function of ${x_\mu}$ only. Applying this to the right-hand-side of \eqref{sccurcond} we 
obtain the condition
\begin{equation}\label{EMfcond}
    \frac{\partial^{2n-2}}{\partial x_\mu^{2n-2}} \bigl(U_\mu\,f^2\bigr)\quad \text{must be a function of ${x_\mu}$ only.}
\end{equation}
It was checked by \textit{Mathematica} that this conditions does not hold 
for the electromagnetic field given by \eqref{EMfres}, at least
for the lowest non-trivial values of ${n}$.
It seems that the main problem is that ${\sccur}$ behaves as 
 ${\sum h_\mu/U_\mu}$ while ${f^2}$ as a square of such sums.

We thus may conclude that in a generic even dimension the electromagnetic field of the
form \eqref{EMFansatz}, \eqref{EMfres} cannot
couple to the metric given by \eqref{metricE}.

\section{Odd dimensions, ${D=2n+1}$}

\subsection*{The metric}

Let us briefly review modifications which appear in the odd dimensional case.
We have an additional coordinate ${\psi_n}$ which labels an \vague{unpaired} angular direction. 
The metric contains an additional term,
\begin{equation}\label{metricO}
\begin{split}
\tens{g}&=\sum_{\mu=1}^n\;\biggl[\; \frac{U_\mu}{X_\mu}\,{\grad x_{\mu}^{\;\,2}}
  +\, \frac{X_\mu}{U_\mu}\,\Bigl(\,\sum_{k=0}^{n-1} \A{k}_{\mu}\grad\psi_k \Bigr)^{\!2} \;\biggr]\\
  &\mspace{190mu}+\frac{c}{\A{n}}\Bigl(\sum_{k=0}^n \A{k}\grad\psi_k\!\Bigr)^{\!2} 
  \period\end{split}
\end{equation}
Here ${c}$ is a conventional constant.\footnote{%
The constant ${c}$ could be eliminated by a appropriate rescaling of the coordinates and 
other parameters.}
We can again introduce the frames of 1-forms given by \eqref{eforms}
completed with 1-forms which we label by the index ${{\hat0}\equiv 2n+1}$
\begin{equation}\label{eframeO}
\ep^{\hat0} = \sum_{k=0}^{n}\A{k}\grad\psi_k\comma
\eb^{\hat0} = {\textstyle\bigl[\frac{c}{\A{n}}\bigr]^{\!1\!/2}}\,\ep^{\hat0}\period
\end{equation}
The dual frames then become
\begin{equation}\label{eframevecO}
\begin{aligned}
\ep_{\hat\mu} &= \sum_{k=0}^{n}\frac{(-x_{\mu}^2)^{n\!-\!1\!-\!k}}{U_{\mu}}\,\cv{\psi_{k}}\commae&
\eb_{\hat\mu} &= {\textstyle\bigl[\frac{U_\mu}{X_\mu}\bigr]^{\!1\!/2}}\,\ep_{\hat\mu}\commae\\
\ep_{\hat0} &= \frac1{\A{n}}\cv{\psi_n}\commae&
\eb_{\hat0} &= {\textstyle\bigl[\frac{\A{n}}{c}\bigr]^{\!1\!/2}}\,\ep_{\hat0}\period
\end{aligned}
\end{equation}
The metric diagonalizes
\begin{equation}
\tens{g}
     = \sum_{\mu=1}^n\,
          \biggl(\,\frac{U_\mu}{X_\mu}\,\ep^\mu \ep^\mu 
          + \frac{X_\mu}{U_\mu}\,\ep^{\hat \mu} \ep^{\hat \mu}\,\biggr)
          +\frac{c}{\A{n}}\ep^{\hat0}\ep^{\hat0}
     = \sum_{a=1}^D \eb^a \eb^a\commae
\end{equation}
as well as the the Ricci tensor \cite{HamamotoEtal:2007}
\begin{equation}\label{RicciO}
  \Ric = -\sum_{\mu=1}^n \; r_\mu\; \bigl(\eb^{\mu}\eb^{\mu}+\eb^{\hat\mu}\eb^{\hat\mu}\bigl)
        \;-\; r_0\; \eb^{\hat0}\eb^{\hat0}\period
\end{equation}
Here
\begin{equation}
\begin{split}
  r_\mu 
    &= \frac12\,\frac{\bX_\mu''}{U_\mu}
      +\frac1{2x_\mu}\,\frac{\bX_\mu'}{U_\mu}
      +\sum_{\substack{\nu=1\\\nu\neq\mu}}^n\frac1{U_\nu}\,\frac{x_\nu \bX_\nu' \!-\! x_\mu \bX_\mu'}{x_\nu^2\!-\!x_\mu^2}\\
    &= \frac\partial{\partial x_\mu^2}\Biggl[\,\sum_{\nu=1}^{n}\frac{x_\nu\bX_\nu'}{U_\nu}\Biggr]\commae\\
  r_0 &= \sum_{\nu=1}^n\frac{\bX_\nu'}{x_\nu U_\nu}\commae
\end{split}
\end{equation}
and we used shifted metric functions
\begin{equation}\label{shiftedX}
  \bX_\mu = X_\mu + \frac{c}{x_\mu^2}\period
\end{equation}
The scalar curvature becomes
\begin{equation}\label{sccurO}
  \sccur = -\sum_{\nu=1}^n \; \frac{\bX_\nu''}{U_\nu} 
          - 2\sum_{\nu=1}^n \; \frac1{x_\nu}\frac{\bX_\nu'}{U_\nu} \period
\end{equation}
Finally, the vacuum Einstein equations require
\begin{equation}\label{BHXsO}
  X_\mu = -\frac{c}{x_\mu^2} + b_\mu + \sum_{k=0}^{n-1}\, c_{k}\, x_\mu^{2k}\commae
\end{equation}

The spacetime has ${n+1}$ Killing vectors ${\cv{\psi_j}}$
and there exists the principal Killing--Yano tensor
dual to the conformal Killing--Yano tensor which is given again by 
\eqref{CCKY}.

\subsection*{Algebraically special test electromagnetic field}

As in the even dimensions we look for the electromagnetic field
with the structure given by \eqref{EMFansatz}.
On could ask if in the odd dimension couldn't be
this ansatz extended by an additional term 
related to the \vague{unpaired} direction,
i.e., by the term of the form ${f_0\, \grad x\wedge\ep^{\hat0}}$
with functions ${f_0}$ and ${x}$ independent of the coordinates ${\psi_j}$.
However, we do not consider such terms since
they lead to some unwanted consequences (e.g., they lead to a non-diagonal 
stress-energy tensor).

It follows from the first Maxwell equations that the vector potential
can be written as
\begin{equation}\label{EMAO}
  \tens{A} = \frac{g_0}{\A{n}}\ep^{\hat0} 
     + \sum_{\mu=1}^n \frac{g_\mu }{U_\mu}\;\ep^{\hat\mu}\commae
\end{equation}
where ${g_\mu}$ is a function of ${x_\mu}$ only and ${g_0}$ is a constant.
The components of the Maxwell tensor can be again generated by \eqref{EMfphi} from
the auxiliary potential
\begin{equation}\label{EMphiO}
    \phi = \frac{g_0}{\A{n}} + \sum_{\nu=1}^n \frac{g_\nu}{U_\nu} \commae
\end{equation}
and in terms of ${g}$'s they take the form
\begin{equation}\label{EMfO}
  f_\mu =  \frac{g_\mu'}{U_\mu}
  +2\,x_\mu\,\sum_{\substack{\nu=1\\\nu\neq\mu}}^n\frac{1}{U_\nu}\,\frac{\,g_\nu - \,g_\mu}{x_\nu^2-x_\mu^2}
  -\frac{2g_0}{x_\mu\A{n}}\period
\end{equation}
The electromagnetic current becomes
\begin{equation}\label{EMJO}
  \tens{J} = -2 \sum_{\mu=1}^n\, \frac{\partial}{\partial x_\mu^2}\biggl[
      x_\mu^2\sum_{\nu=1}^n x_\mu^{-1}\phi_\nu\biggr]\ep_{\hat\mu}
      +2\sum_{\nu=1}^n\frac1{x_\mu}\,\phi_{,\nu}\ep_{\hat0}\period
\end{equation}
Solving the second Maxwell equation without sources (i.e., ${\tens{J}=0}$)
we obtain
\begin{equation}\label{EMggenO}
  g_\mu =  e_\mu -\frac{g_0}{x_\mu^2}+ \sum_{k=0}^{n-2}\, a_k\, \bigl(-x_\mu^2\bigr)^{n-1-k}\period
\end{equation}
The last two terms are gauge trivial and
the vector potential can thus be written as
\begin{equation}\label{EMAresO}
  \tens{A} =  \sum_{\mu=1}^n \frac{e_\mu }{U_\mu}\;\ep^{\hat\mu}\commae
\end{equation}
and the auxiliary potential ${\phi}$ as
\begin{equation}\label{EMphiresO}
    \phi = \sum_{\nu=1}^n \frac{e_\nu}{U_\nu} \period
\end{equation}
Finally, for the components of the Maxwell tensor we find
\begin{equation}\label{EMfresO}
  f_\mu = 2\,x_\mu\,\sum_{\nu=1}^n\frac{1}{U_\nu}\,\frac{e_\nu - e_\mu}{x_\nu^2-x_\mu^2}\period
\end{equation}

\subsection*{The Chern--Simons term}

In an odd number of dimensions we can consider Chern--Simons modification of the 
electromagnetic field. Since the Chern-Simons term in the action
does not refer to the metric, it does not change the Einstein equation
and the stress-energy tensor of the electromagnetic field.
However, it modifies the divergence ${-\covd\cdot\tens{F}}$ in the second Maxwell equation by adding a
non-linear term ${\tens{J}_{\mathrm{ChS}}}$ proportional (with a constant coefficient) to
the Hodge dual of the ${n}$-th wedge power of ${\tens{F}}$,
\begin{equation}
  \tens{J}_{\mathrm{ChS}}\propto*\, \bigl(\underset{n\text{ times}}{\underbrace{\tens{F}\wedge\dots\wedge\tens{F}}}\bigr)\period
\end{equation}
Since the Levi-Civita tensor ${\tens{\eps}}$ used in the Hodge dual
is given by the product ${\eb^{1}\wedge\dots\wedge\eb^{2n+1}}$
we find that the Chern--Simons term ${\tens{J}_{\mathrm{ChS}}}$ for the field \eqref{EMFansatz}
is aligned with the \vague{unpaired} direction
\begin{equation}\label{EMJChS}
  \tens{J}_{\mathrm{ChS}}\propto  \biggl[\prod_{\mu=1}^n f_\mu\biggr]\; \eb_{\hat0}\period
\end{equation}
We thus have to solve the Maxwell equation ${\tens{J}=  \tens{J}_{\mathrm{ChS}}}$
with ${\tens{J}}$ given by \eqref{EMJO}. The condition that ${\tens{J}}$
vanishes in ${\eb_{\hat\mu}}$, ${\mu=1,\dots,n}$, directions is only slightly
weaker than that we solved in the previous section. We find that
the vector potential ${\tens{A}}$, the auxiliary potential ${\phi}$, 
and the components of the electromagnetic field are
\begin{align}
  \tens{A} &=  \sum_{\mu=1}^n \frac{e_\mu + a\log x_\mu^2 }{U_\mu}\;\ep^{\hat\mu}\comma
  \phi = \sum_{\nu=1}^n \frac{e_\nu + a\log x_\mu^2 }{U_\nu} \commae\notag
\\\label{EMresChS}
  f_\mu &= 2\,x_\mu\,\sum_{\nu=1}^n\frac{1}{U_\nu}\,\frac{e_\nu - e_\mu}{x_\nu^2-x_\mu^2}\\\notag
        &\;\;+2 a \,\biggl[\frac1{x_\mu}+
         x_\mu\!\sum_{\substack{\nu=1\\\nu\neq\mu}}^n\frac1{U_\nu}\,\frac{\log x_\nu^2-\log x_\mu^2}{x_\nu^2-x_\mu^2}\biggr]
  \commae
\end{align}
where ${a}$ is a constant parameter. Moreover, the source term ${\tens{J}}$
must be
\begin{equation}
\tens{J} = \frac{4a}{\A{n}}\,\eb^{\hat0}\period
\end{equation}
Although we did not prove it rigorously it seems evident that 
this component cannot be equal to \eqref{EMJChS} with ${f_\mu}$
given by \eqref{EMresChS}. We thus conclude that the Chern--Simons 
modification of the algebraically special electromagnetic field is
not possible.

\subsection*{Einstein--Maxwell equations in odd ${D}$}

The stress-energy tensor for the field \eqref{EMFansatz}
in odd dimensions is also diagonal 
(with the additional component ${8\pi T_{\hat0\hat0}=-f^2}$).
The condition for the scalar curvature following
from the Einstein equations is again
given by \eqref{sccurcond}, and 
by the same argument as in the even dimensions 
(now with help of \eqref{sccurO}) we conclude 
that the algebraically special electromagnetic field \eqref{EMFansatz}, \eqref{EMfresO}
cannot couple to the metric given by \eqref{metricO}.

\section{Summary}

We have found explicitly
the test electromagnetic field on the background
of the high-dimensional rotating NUT charged black hole 
\cite{ChenLuPope:2006}.
The field solves the Maxwell equations even in a broader 
class of spacetimes since the specific form of the
metric functions ${X_\mu}$ is not needed.
The common feature of these spacetimes is the presence
of ${\lceil D/2\rceil}$ Killing vectors and the existence
of the principal Killing--Yano tensor \cite{KrtousEtal:2007a}.
The constructed electromagnetic field is adjusted to this
structure -- it shares the explicit symmetries and it has
the same eigenspaces as the principal conformal Killing--Yano tensor.

The electromagnetic field depends on ${n=\lfloor D/2\rfloor}$
constants ${e_\mu}$ related to the global electric and magnetic charges.
It generalizes the field known on the background
of the Pleba\'nski--Demia\'nski spacetime in ${D=4}$ dimensions.
In this case the metric functions can be modified in such
a way that the field and the metric solve the full Einstein--Maxwell equations.
Unfortunately, an analogous modification is not possible in a generic dimension.

Finally, we have shown that the Chern--Simons generalization in an odd dimmension
is also not permitted for the electromagnetic field of this form.

\section*{Acknowledgments}
This work was supported by the grant GA\v{C}R 202/06/0041 and 
by the Czech Ministry of Education under the project LC06014.
The author would like to thank Don N. Page, Valeri P. Frolov, David
Kubiz\v{n}\'ak and Ji\v{r}\'{\i} Podolsk\'y for numerous discussions,
and Ji\v{r}\'{\i} Podolsk\'y and David Kubiz\v{n}\'ak 
for reading the manuscript.

\appendix

\section{}

The functions ${\A{k}_\mu}$ and ${U_\mu}$ defined in \eqref{UAdef}
satisfy the following useful identities
\begin{gather}
  \sum_{\mu=1}^n \A{i}_\mu \frac{(-x_\mu^2)^{n\!-\!1\!-\!j}}{U_\mu} = \delta^i_j\comma
  \sum_{j=0}^{n-1} \A{j}_\mu \frac{(-x_\nu^2)^{n\!-\!1\!-\!j}}{U_\nu} = \delta^\nu_\mu\commae  \notag
  \\[-1ex]\label{AUrel}
  \\[-1ex]\notag
  \sum_{\mu=1}^n \A{i}_\mu \frac{(-x_\mu^2)^{n}}{U_\mu} = -\A{j+1}\comma
  \sum_{\mu=1}^n \frac{\A{j}_\mu}{x_\mu^2U_\mu} = \frac{\A{j}}{\A{n}}\commae
\end{gather}
\begin{gather}
  \sum_{j=0}^n (n\!-\!j)\, \A{j} \frac{(-x_\mu^2)^{n\!-\!1\!-\!j}}{U_\mu} = 1\commae\notag\\
  \sum_{j=0}^{n-1} (n\!-\!1\!-\!j)\, \A{j}_\mu \frac{(-x_\mu^2)^{n\!-\!1\!-\!j}}{U_\mu} 
    = \sum_{\substack{\nu=1\\\nu\neq\mu}}^{n}\frac{x_\mu^2}{x_\mu^2-x_\nu^2}\commae\label{sumn1jAxn1j}\\
  \sum_{j=0}^{n-1} (n\!-\!1\!-\!j)\, \A{j}_\nu \frac{(-x_\mu^2)^{n\!-\!1\!-\!j}}{U_\mu} 
    = \frac{x_\mu^2}{x_\mu^2-x_\nu^2}\quad\text{for}\;\mu\neq\nu\commae\notag
\end{gather}
and 
\begin{equation}
  \frac1{U_\mu}\sum_{\substack{\nu=1\\\nu\neq\mu}}^n\frac1{x_\nu^2-x_\mu^2}
    = - \sum_{\substack{\nu=1\\\nu\neq\mu}}^n \frac1{U_\nu}\,\frac1{x_\nu^2-x_\mu^2}\period
\end{equation}

We list also external derivatives of the function ${1/U_\mu}$ and of the 1-form ${\ep^{\hat\mu}}$
which have been used repeatedly in the computations:
\begin{equation}\label{d1U}
  \grad\frac1{U_\mu} = -\frac1{U_\mu}\sum_{\substack{\nu=1\\\nu\neq\mu}}^n\frac2{x_\nu^2-x_\mu^2}
    \bigl(x_\nu\,\ep^{\nu} -x_\mu\,\ep^{\mu}\bigr)\commae
\end{equation}
\begin{equation}\label{deps}
  \grad\ep^{\hat\mu}=\sum_{\substack{\nu=1\\\nu\neq\mu}}^n \frac2{x_\nu^2\!-\!x_\mu^2}
  \bigl(x_\nu\,\ep^\nu\wedge\ep^{\hat\mu}-x_\nu\,\ep^\nu\wedge\ep^{\hat\nu}\bigr)\period
\end{equation}

Finally, let us formulate explicitly an important lemma
(which has been already used implicitly in \cite{HamamotoEtal:2007})
concerning properties of the sums ${\sum_\nu h_\nu/U_\nu}$.
Let's consider the equation
\begin{subequations}\label{sumhU}
\begin{equation}\label{sumhUa}
  \sum_{\mu=1}^n\frac{h_\mu}{U_\mu}=0\commae
\end{equation}
where ${h_\mu}$ are functions of a single variable ${x_\mu}$ only. 
Then these functions are given by a single polynomial
\begin{equation}\label{sumhUb}
  h_\mu = \sum_{k=0}^{n-2} \,c_k\, x_\mu^{2k}
\end{equation}
with arbitrary coefficients ${c_k}$, ${k=0,\dots,n\!-\!2}$.

The fact that the polynomial functions \eqref{sumhUb} solve equation \eqref{sumhUa}
follows from \eqref{AUrel}. The opposite implication is less 
trivial---first one has to show that 
${h_\mu}$ must be polynomials of the order ${n-2}$
(by differentiating repeatedly ${U_\mu\sum_\nu h_\nu/U_\nu}$).
Next, a more intricate task is to prove that the polynomials for 
different ${\mu}$ are the same. It can be achieved by an induction in~${n}$.

The solution of the functional equation \eqref{sumhUa} with a non-trivial right-hand-side
is then given by the sum of a particular solution 
with the homogeneous solution \eqref{sumhUb}.
We mention three important particular solutions for simple right-hand-side terms:
\begin{equation}
\begin{gathered}
  \sum_{\mu=1}^n \frac{(-x_\mu^2)^{n-1}}{U_\mu} = 1\commae\\
  \sum_{\mu=1}^n \frac{(-x_\mu^2)^n}{U_\mu} = -\A{1}\commae\\
  \sum_{\mu=1}^n \frac{1}{x_\mu^2 U_\mu} = \frac1{\A{n}}\commae
\end{gathered}
\end{equation}
\end{subequations}
all following from the relations \eqref{AUrel}.

The property \eqref{sumhU} has been used, for example, to derive
the specific forms \eqref{BHXsE} and  \eqref{BHXsO} of the metric 
functions ${X_\mu}$ starting from the condition ${\sccur=\text{constant}}$ 
with the scalar curvature given by \eqref{sccurE} or \eqref{sccurO}.
It has been used also to find the components \eqref{EMggenE}
and \eqref{EMggenO} of the vector potential.


\begin{thebibliography}{10}

\bibitem{Tangherlini:1963}
F.~R. Tangherlini, Schwartzschild Field in $N$ Dimensions and the
  Dimensionality of Space Problem, Nuovo Cimento {\bf 27},  3  (1963).

\bibitem{MyersPerry:1986}
R.~C. Myers and M.~J. Perry, Black holes in higher dimensional space-times,
  Ann. Phys. (N.Y.) {\bf 172},  304  (1986).

\bibitem{HawkingEtal:1999}
S.~W. Hawking, C.~J. Hunter, and M.~M. Taylor-Robinson, Rotation and the
  AdS/CFT correspondence, Phys. Rev. D {\bf 59},  064005  (1999),
  arXiv:hep-th/9811056.

\bibitem{GibbonsEtal:2004}
G.~W. Gibbons, H. L\"u, D.~N. Page, and C.~N. Pope, Rotating Black Holes in
  Higher Dimensions with a Cosmological Constant, Phys. Rev. Lett. {\bf 93},
  171102  (2004), arXiv:hep-th/0409155.

\bibitem{GibbonsEtal:2005}
G.~W. Gibbons, H. L\"u, D.~N. Page, and C.~N. Pope, The General Kerr-de Sitter
  Metrics in All Dimensions, J. Geom. Phys. {\bf 53},  49  (2005),
  arXiv:hep-th/0404008.

\bibitem{ChenLuPope:2006}
W. Chen, H. L\"u, and C.~N. Pope, General Kerr-NUT-AdS metrics in all dimensions,
  Class. Quantum Grav. {\bf 23},  5323  (2006), arXiv:hep-th/0604125.

\bibitem{HamamotoEtal:2007}
N. Hamamoto, T. Houri, T. Oota, and Y. Yasui, Kerr-NUT-de Sitter curvature in
  all dimensions, J. Phys. {\bf A40},  F177  (2007), arXiv:hep-th/0611285.

\bibitem{KubiznakFrolov:2007}
D. Kubiz\v{n}\'ak and V.~P. Frolov, Hidden symmetry of higher dimensional
  Kerr-NUT-AdS spacetimes, Class. Quantum Grav. {\bf 24},  F1  (2007),
  arXiv:gr-qc/0610144.

\bibitem{KrtousEtal:2007a}
P. Krtou\v{s}, D. Kubiz\v{n}\'ak, D.~N. Page, and V.~P. Frolov, Killing-Yano
  tensors, rank-2 Killing tensors, and conserved quantities in higher
  dimensions, J. High Energy Phys. JHEP02(2007)004,
  arXiv:hep-th/0612029.

\bibitem{PageEtal:2007}
D.~N. Page, D. Kubiz\v{n}\'ak, M. Vasudevan, and P. Krtous, Integrability of
  geodesic motion in general Kerr-NUT-AdS spacetimes, Phys. Rev. Lett. {\bf
  98},  061102  (2007), arXiv:hep-th/0611083.

\bibitem{KrtousEtal:2007b}
P. Krtou\v{s}, D. Kubiz\v{n}\'ak, D.~N. Page, and M. Vasudevan, Constants of
  Geodesic Motion in Higher-Dimensional Black-Hole Spacetimes, 
  arXiv:0707.0001~[hep-th].

\bibitem{FrolovEtal:2007}
V.~P. Frolov, P. Krtou\v{s}, and D. Kubiz\v{n}\'ak, Separability of
  Hamilton-Jacobi and Klein-Gordon equations in general Kerr-NUT-AdS
  spacetimes, J. High Energy Phys. JHEP02(2007)005,
  arXiv:hep-th/0611245.

\bibitem{AlievFrolov:2004}
A.~N. Aliev and V.~P. Frolov, Five dimensional rotating black hole in a uniform
  magnetic field: The gyromagnetic ratio, Phys. Rev. D {\bf 69},  084022
  (2004), arXiv:hep-th/0401095.

\bibitem{Aliev:2005}
A.~N. Aliev, Charged Slowly Rotating Black Holes in Five Dimensions, Mod. Phys.
  Lett. {\bf A21},  751  (2006), arXiv:gr-qc/0505003.

\bibitem{Aliev:2006}
A.~N. Aliev, Rotating black holes in higher dimensional Einstein-Maxwell
  gravity, Phys. Rev. D {\bf 74},  024011  (2006), arXiv:hep-th/0604207.

\bibitem{Aliev:2007}
A.~N. Aliev, Electromagnetic Properties of Kerr-Anti-de Sitter Black Holes,
  Phys. Rev. D {\bf 75},  084041  (2007), arXiv:hep-th/0702129.

\bibitem{BrihayeDelsate:2007}
Y. Brihaye and T. Delsate, Charged-rotating black holes and black strings in
  higher dimensional Einstein-Maxwell theory with a positive cosmological
  constant, 2007, arXiv:gr-qc/0703146.

\bibitem{KunzEtal:2007}
J. Kunz, F. Navarro-Lerida, and E. Radu, Higher dimensional rotating black
  holes in Einstein-Maxwell theory with negative cosmological constant, Phys.
  Lett. {\bf B649},  463  (2007), arXiv:gr-qc/0702086.

\bibitem{ChenLu:2007}
W. Chen and H. L\"u, Kerr-Schild Structure and Harmonic 2-forms on
  (A)dS--Kerr--NUT Metrics, 2007, arXiv:0705.4471 [hep-th].

\bibitem{Carter:1968}
B. Carter, Hamilton-Jacobi and Schrodinger Separable Solutions of Einstein's
  Equations, Commun. Math. Phys. {\bf 10},  280  (1968).

\bibitem{PlebanskiDemianski:1976}
J. Pleba\'{n}ski and M. Demia\'{n}ski, Rotating charged and uniformly
  accelerated mass in general relativity, Ann. Phys. (N.Y.) {\bf 98},  98
  (1976).

\bibitem{GriffithsPodolsky:2005}
J.~B. Griffiths and J. Podolsk\'y, Accelerating and rotating black holes, Class.
  Quantum Grav. {\bf 22},  3467  (2005), arXiv:gr-qc/0507021.

\bibitem{GriffithsPodolsky:2006a}
J.~B. Griffiths and J. Podolsk\'y, Global aspects of accelerating and rotating
  black hole space-times, Class. Quantum Grav. {\bf 23},  555  (2006),
  arXiv:gr-qc/0511122.

\bibitem{GriffithsPodolsky:2006b}
J.~B. Griffiths and J. Podolsk\'y, A new look at the Plebanski-Demianski family
  of solutions, Int. J. Mod. Phys. {\bf D15},  335  (2006),
  arXiv:gr-qc/0511091.

\bibitem{PodolskyGriffiths:2006}
J. Podolsk\'y and J.~B. Griffiths, Accelerating Kerr-Newman black holes in
  (anti-)de Sitter space-time, Phys. Rev. D {\bf 73},  044018  (2006),
  arXiv:gr-qc/0601130.

\bibitem{GriffithsPodolsky:2007}
J.~B. Griffiths and J. Podolsk\'y, On the parameters of the Kerr-NUT-(anti-)de
  Sitter space-time, Class. Quantum Grav. {\bf 24},  1687  (2007),
  arXiv:gr-qc/0702042.

\end{thebibliography}

\pagebreak

\end{document}